\documentclass[preprint,12pt]{elsarticle}
\usepackage{amssymb}
\usepackage{color}
\usepackage{epsfig}
\usepackage{amsmath}

\begin{document}
\begin{frontmatter}

\title{Heterogeneous noise enhances spatial reciprocity}

\author[a]{Yao Yao}
\author[a]{Shen-Shen Chen}

\address[a]{Henan Electric Power Survey \& Design Institute, Zhengzhou, He Nan, 450007,  China\\
{E-mail: yaoyaohust01@gmail.com (YY); css0703@126.com(SSC)}}

\begin{abstract}
Recent research has identified the heterogeneity as crucial for the evolution of cooperation in spatial population. However, the influence of heterogeneous noise is still lack. Inspired by this interesting question, in this work, we try to incorporate heterogeneous noise into the evaluation of utility, where only a proportion of population possesses noise, whose range can also be tuned. We find that increasing heterogeneous noise monotonously promotes cooperation and even translates the full defection phase (of the homogeneous version) into the complete cooperation phase. Moreover, the promotion effect of this mechanism can be attributed to the leading role of cooperators who have the heterogeneous noise. These type of cooperators can attract more agents penetrating into the robust cooperator clusters, which is beyond the text of traditional spatial reciprocity. We hope that our work may shed light on the understanding of the cooperative behavior in the society.
\end{abstract}

\begin{keyword}
Evolutionary Game, Spatial Reciprocity, Noise, Heterogeneity, Network Dynamics
\end{keyword}

\end{frontmatter}

\section{Introduction}
Understanding the emergency of cooperation among the population of selfish individuals denotes one challenge in natural and social science \cite{Axelrod84,Smith95}. To interpret the survival of cooperation, a theoretical framework that has shed light onto this long-standing issue is the evolutionary game theory \cite{Nowak06,PS10Biosystem,WZJTB,PGSFM13INTERFACE}. In particular, a simple, paradigmatic model, the prisoner¡¯s dilemma (PD) game, has attracted much attention, both theoretical and experimental \cite{VA01PRE,SKTHI12EPL,STWKHI12PRE,RLW07PRE,GFRTCSM12PNAS,WSP12SR,SWP12SR2}. In its basic version, two players simultaneously decide to
adopt one of two strategies: cooperation (C) and defection (D). They will receive the reward $R$ if both cooperate, and the punishment $P$ if both defect. However, if one player defects while the other decides to cooperate, the former will get the temptation $T$ while the latter will get the sucker's payoff $S$. These payoffs satisfy the ranking $T>R>P>S$ and $2R>T+S$; thus, defection optimizes the individual payoff, in spite of the fact that mutual cooperation could yield a higher collective benefit. If no special mechanism is introduced into the population, resulting is a social dilemma, which leads to widespread defection.

During the past decades, a great number of scenarios has been proposed that can overcome this dilemma and lead to the evolution of cooperation \cite{vo1,vo2,vo3,vo4,vo5,vo6}. Whereas, Nowak attributed all these to five mechanisms: kin selection, direct reciprocity, indirect reciprocity, network reciprocity, and group selection \cite{Nowak06Science}. Among the five mechanisms, network reciprocity, where players are arranged on the spatially structured topology and interact only with their direct neighbors, has attracted the greatest interest (for comprehensive reviews refer to Ref. \cite{GF07PR}), because cooperators can survive by means of forming compact clusters, which minimize the exploitation by defectors and protect those cooperators that are located in the interior of such clusters. In line with this achievement, the role of spatial structure, and its various underlying promoting mechanisms, in evolutionary games
have been intensively explored. Examples include heterogeneous activity \cite{he1,he2,he3}, complex networks \cite{com1,com2,com3},  different rewiring mechanisms \cite{rew1,rew2,rew3,rew4}, reward and punishment mechanism \cite{re1,re2,re3,re4,re5,re6}, environment influence \cite{en1,en2}, differences in evolutionary time scales \cite{time1,time2}, interdependent topology \cite{int1,int2,int3}, the impact of reputation \cite{rep1,rep2}, to name but a few.  Looking at some examples more specifically, in a recent research framework \cite{diu1,diu2}, where players were arranged on the diluted spatial networks, an optimal state similar to the percolation theory could maximize cooperation. In \cite{exp1,exp2} it was shown that the terminologies of enduring (END) and expanding (EXP) periods could bright light to the understanding of the so-called "spatial reciprocity", even if the conditions did not necessarily favor the spreading of cooperators.

Except for the above scenarios, another mechanism that attracts great attention is the influence of noise on the evolution of cooperation. We can examine some achievements up to now. When mapping the heterogeneous strategy transfer capability to age structure, the robust promotion of cooperation was reported \cite{noise1}. Moreover, the heterogeneity diversity of players allowed for cooperative behavior to prevail even if the temptations to defect were large \cite{he3}. Recently, it was shown if noise was involved into the payoffs of different strategies, cooperation was largely enhanced as well \cite{noise2}. Although it is undisputable that noise is a relevant ingredient when modeling a population, there are, however, several aspects which greatly differentiates from realistic situations. One situation of particular relevance that has received relatively little is the case of heterogeneous noise, where some people inherently poss the noise while this influence is congenitally eradicated for some more. Inspired by this fact, one interesting question appears: if we introduce heterogeneous noise into game model, is it beneficial for the evolution of cooperation?

To unveil the answer of this puzzle, we thus study the prisoner's dilemma game with the
consideration of heterogonous noise that maps to players's fitness. Comparing with the works of influence of merely noise, a proportion of population is assumed to have noise and keep constant during the whole evolution process.  By means of numerical simulations we demonstrate, compared with previous reports, that this simple mechanism can actually promote the evolution of cooperation further. While the promotion trait is leaded by the agents who have heterogeneous noise. In the remainder of this paper we will first describe
the considered evolutionary game, subsequently present the main results, and summarize our conclusions finally.

\section{Evolution Game Model}
We consider an evolutionary prisoner's dilemma game that is characterized with the temptation to defect $T$ (the highest payoff received by a defector if playing against a cooperator), reward for mutual cooperation $R=1$, and both the punishment for mutual defection $P$ as well as the sucker's payoff $S$ (the lowest payoff received by a cooperator if playing against a defector) equaling 0. As a standard practice, $1<T\leq2$ ensures a proper payoff ranking ($T>R>P \ge S$) and captures the essential social dilemma between individual and common interests \cite{nowak_n92b}. It is worth mentioning that though we choose a simply and weak version (namely, $S=0$), our observations are robust and can be observed in the full parameterized space as well \cite{santos_pnas06}.

Throughout this work, each player $x$ is initially designated either as a cooperator (C) or defector (D) with equal probability. Two types of players are distinguished and their spatial distribution is described by an Ising formalism ($n_x=A$ or $B$). The portion of players $A$ ($B$) is $\nu$ ($1-\nu$) and keeps constant during the whole simulation process. With respect to the interaction network, we choose the $L \times L$ regular lattice with four nearest neighbors and periodic boundary conditions. The game is iterated forward in accordance with the Monte Carlo simulation procedure comprising the following elementary steps. First, a randomly selected player $x$ evaluates his utility $U_x$ based on the acquired payoff $P_x$ by playing the game with its nearest neighbors. Then, it chooses at random one neighbor $y$ who also gets his utility $U_y$ in the same way and adopts the strategy $s_y$ from the selected player $y$ with the probability
\begin{equation}
W(s_y \to s_x)=\frac{1}{1+exp[(U_x-U_y)/K]},
\end{equation}
where $K$ denotes the uncertainties, or intensity of selection \cite{k1,k2}. Since the effect of $K$ has been extensively investigated \cite{k1,prev}, we simply fix the value of $K$ to be $K=0.1$ in the present work. With regard to the utility $U_x$, it can be defined in the following way
\begin{equation}
U_x=F_x \times \epsilon,
\end{equation}
where reflect the influence of noise, which introduces the noise into the utility, similar to previous treatment \cite{k1}.  Importantly, if player $x$ belongs to $A$ type, it will be assigned one uniformly distributed random number $\Lambda$ in the interval [-1,1]. In this case, we can get the value of $\epsilon=\eta \Lambda$, where $\eta$ is used to control the range of random number. On the contrary, given that player $x$ is type $B$, $\epsilon$ equals to 0 (namely, there is no the influence of noise). Obviously, when $\eta=0$ (or $\nu=0$), it will turn to the traditionally homogeneous case \cite{k1}, while positive values incorporates heterogeneous noise.  During one full Monte Carlo step (MCS) each player has a chance to adopt one of the neighboring strategies once on average.

Results of Monte Carlo simulations presented below were obtained on $200\times 200$ lattices. Key quantity the fraction of cooperators $\rho_C$ was determined within the last $10^4$ full MCS over the total $2\times10^5$ steps. Moreover, since the heterogeneous distribution may introduce additional disturbances, the final results were averaged over up to 100 independent realizations for each set of parameter values in order to assure suitable accuracy.

\section{Results and Analysis}

We start by examine how cooperation trait varies under the influence of heterogeneous noise. Figure 1 features the color map encoding the final fraction of cooperation $\rho_C$ on the $\nu -\eta$ parameter plane. For $\eta=0$ (or $\nu=0$, which can be regarded as the baseline), it returns the traditional version, where the weight of utility is identical among agents and cooperation vanishes under even small temptation to defection. However, when heterogeneous noise is introduced, we can see that everything changes.  It is obvious that increasing $\eta$ and $\nu$ enhances the maximally required temptation that is needed for cooperator to survive.  What is more, the span of cooperators and defectors coexisting expands too, ultimately leading also to an ever-broader arrival to the pure C phase as increases further. More precisely, the minimally required $b$ of cooperation survival is above 1.0375 for $\eta=\nu=0$, but it becomes possible for the dominance of cooperation at $b=1.08$ (under $\eta>0$ and $\nu>0$). This is certainly impressive and motivating for trying to understand the mechanism behind the remarkable promotion of social cooperation. These results suggest that when heterogeneous noise incorporated into the evaluation of utility plays an important role in facilitating the prosperity of cooperation.  In what follows we will systematically examine the validity of this claim.

It is therefore of interest to proceed with exploring the spatial distribution of strategies under the influence of heterogeneous noise. Results presented in
Fig. 2 hint to the performance of the case, where cooperators and defectors of type $A$ are colored by blue and yellow, while the same classes of type $B$ are marked by green and red. Irrespective of which case, what first attracts our attention is that the change of cooperator clusters, through which cooperators can survive and resist the exploitation of defectors. As evidenced in the left panel (small value of $\nu$ or $\eta$), there is slight distribution of heterogeneous noise, cooperators goes still extinct, thus yielding an exclusive dominance of defectors. With the increment of $\nu$ or $\eta$ (middle panel), a fraction of cooperators can survive on the lattice by means of forming clusters,
thereby protecting themselves against the exploitation by defectors. Moreover, it is worth mentioning that these clusters are usually small and discrete, which to some extent
helps to explain why it is impossible to yield the absolute dominance of cooperators. However, if the heterogeneity is sufficient strong (right panel), one can see that cooperators prevail even reach their undisputed dominance, whereby clustering remains their mechanism of spreading and survivability.  Compared with the left panel, it is obvious that the clusters of cooperators become larger and more compact, which further results in less space left for defectors. More importantly, cooperators of type $A$ ($AC$, blue) are basically located at the centers of clusters, while cooperators of type $B$ ($BC$, green) lie along the boundaries of clusters (see Fig.2 for more details). We argue that cooperators of type $A$ (who posses the noise) play a crucial role in sustaining the large clusters of cooperators. Namely, during the evolution process, cooperators of type $A$ will be chosen more likely as potential strategy donors, which induces more cooperators approaching and surrounding them. Consequently, the initial clusters warranted start mushrooming to depress the invasion of defectors. It is also natural that their followers, i.e., cooperators of type $B$, usually lie along the boundaries. In a sea of cooperators this is practically always these followers rather than defectors trying to penetrate into the clusters. This kind of expansion ultimately results in highly robust clusters of cooperators that goes beyond the observation supported by spatial reciprocity alone \cite{nowak_n92b,k1}.

In order to explain the promotive impact of heterogeneous noise on the evolution of cooperation, it is subsequently instructive to examine time courses of $\rho_C$ for different values of the parameter $\eta$ (or $\nu$, due to the similar evolution tide, we do not repeat it).  Figure 3 features how cooperation evolves under different scenarios. What first attracts great attention is the  early stages of evolutionary process (irrespective of the values of $\eta$), where the performance of defectors is better than that of cooperators. This is given that defectors are, as individuals, more successful than cooperators and will thus be chosen more likely as potential strategy donors (similar to the characteristics of EXP period in recent literatures \cite{exp1,epx2}). Moreover, we also note a fact that the small the value of $\eta$, the faster the decline of cooperation. This should be related with the distribution of noise, which has be proved that broader distribution of heterogeneous noise is beneficial for the extension of cooperation clusters from Fig.2. However, with the evolution proceeding, interesting phenomenon takes place. For small value of $\eta$, the decline of cooperation can not stop till the absolute dominance of defector. While for larger value of $\eta$, it quickly restrains from the exploitation of defectors and is in favor of the prosperity of cooperators. Obviously, larger value of $\eta$ can lead to earlier reverse of cooperators (the advantage of of defectors is weakened largely) and finally reach higher level of cooperation. Among this expansion process, broad distribution of cooperators of type $A$ plays a crucial role: these cooperators attract more individuals to form effective clusters through inducing
the transformation from defectors to cooperators. The clusters built by cooperators are impervious to the lure of becoming defectors and able to recover the space of
defectors, which ultimately results in widespread cooperation. Thus, we argue that the early forming of clusters composed by cooperators of type $A$ is determinative, the existence of which can results in widespread cooperation going beyond the tradition version \cite{nowak_n92b,k1}.

Finally, it remains of interest to validate the role of cooperators type $A$ for the effective expansion of clusters. Figure 4 shows the evolution snapshots of particular state: there initially are two circular cooperator domains. It is visually clear that cooperator domains where there is no noise (green circular domain) will suffer from the impingement of defectors. In this case, cooperators initially belonging to this circle can only maintain their territory for a short time (namely, the green circle will vanish soon and is replaced by defectors). Quite surprisingly, the fate of the circular domain of type $A$ (blue circle) is fully different. The domain is always robust, but more importantly is that it can attract more defectors to becomes cooperators. That is to say, these cooperators are impervious to defector attacks, and at the same time let more agents penetrate into the clusters, which eventually causes cooperation spreading across the whole network. Therefore, this fig provides a direct proof that cooperators possessing heterogeneous noise play a significant role for the effective expansion of cooperation behavior, which can be regarded as a valuable clue to explaining more social dilemmas \cite{soc1,soc2,soc3}.

\section{Summary}

To sum, we have studied the influence of heterogeneous noise on the evolution of cooperation in spatial prisoner's dilemma game. Different from previous works about noise, we allow part of agents to possess noise, which can be incorporated into the evaluation of utility. Through systematic computer simulations, we have found that heterogeneous noise is greatly beneficial for enhancing cooperation level (namely, facilitates the spatial reciprocity). While the essence of this promotion effect can be attributed to the leading role of cooperators who possess noise.  These cooperators not only resist the exploitation of defectors, but also promote the transformation of defectors to cooperators. Except for the time course, we provide a direct observation for this argument as well. Since various types of dilemmas are ubiquitous in our life,  this simple mechanism can supply the clue for more cases. Moreover, we also hope that it can inspire further study for  some social puzzles via a co-evolutionary process \cite{PS10Biosystem} and its application \cite{add1,add2}.

\section*{Acknowledgments}
This project is supported by the National Natural Science
Foundation of China under Grant No. 60904063.


\begin{figure}
\begin{center}
\includegraphics[width=9.5cm]{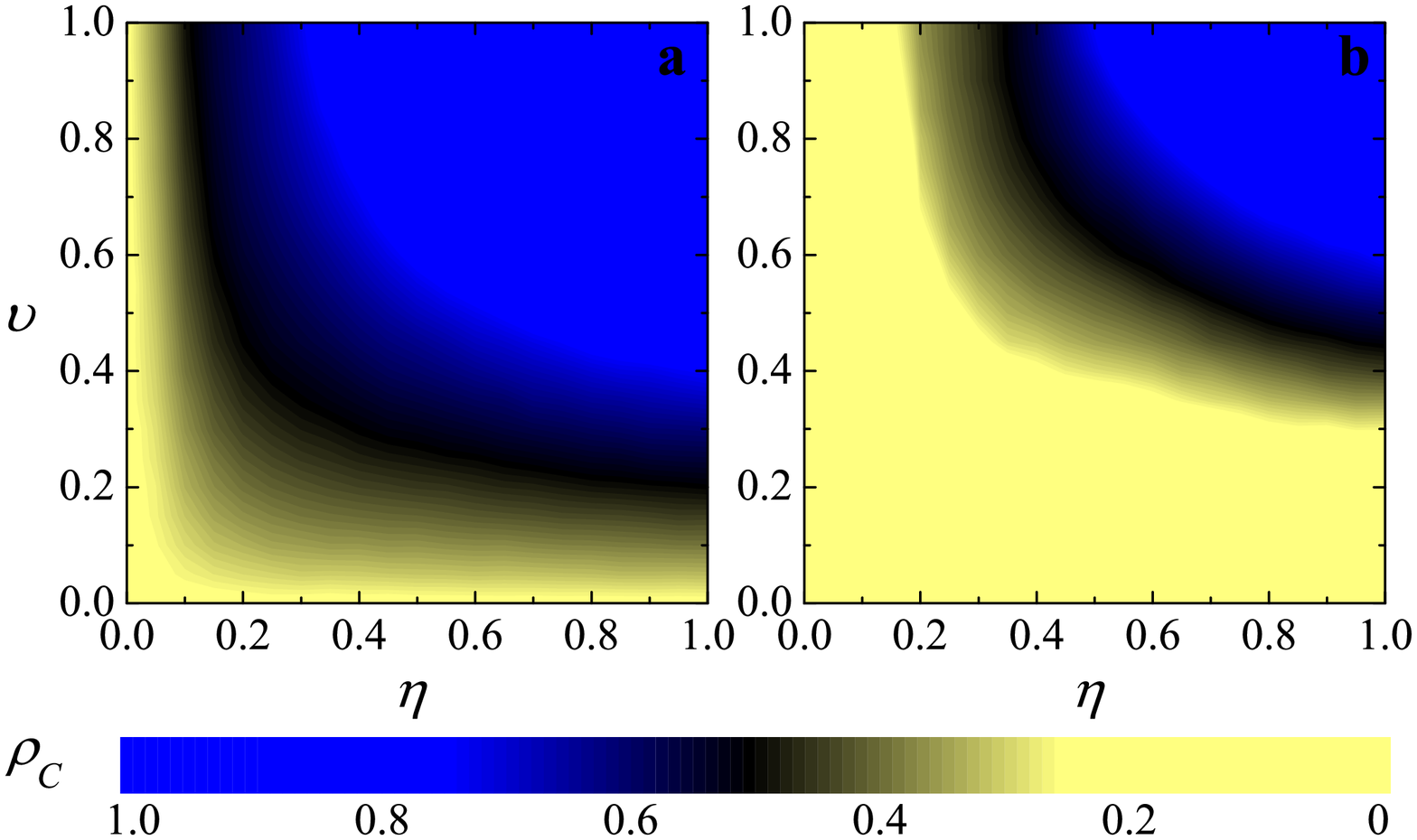}
\caption{Color map encoding the fraction of cooperators $\rho_C$ on the $\nu-\eta$ parameter plane. When $\nu=0$ (or $\eta=0$), it returns to the classical homogeneous version, where cooperation dies out soon. After introducing heterogeneous noise, we can see that cooperation is obviously promoted with the increment of $\nu$ and $\eta$, namely, heterogeneous noise enhances spatial reciprocity. The values of $b$ are 1.03 and 1.08 for panels {\bf a} and {\bf b}, respectively.}
\end{center}
\end{figure}

\begin{figure}
\begin{center}
\includegraphics[width=13cm]{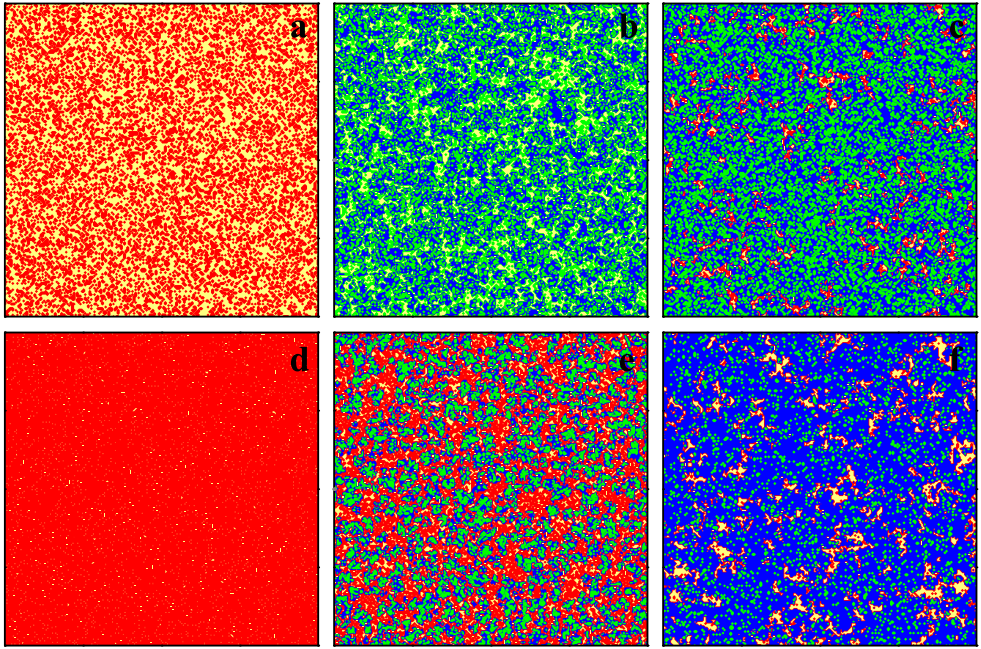}
\caption{Characteristic snapshots of strategy distributions on the square lattice. Top row depicts results for the fraction of type $A$ $\nu=0.7$, while from left ({\bf a}) to right ({\bf c}) the values of $\eta$ are 0.1, 0.5 and 0.9, respectively. Bottom panels shows the results for the parameter value $\eta=0.6$, while from left ({\bf d}) to right ({\bf f}) the values of $\nu$ are 0.1, 0.5 and 0.9. Cooperators of type $A$ ($AC$) and $B$ ($BC$) are colored blue and green, respectively. Defectors of type $A$ ($AD$) and $B$ ($BD$), on the other hand, are colored yellow and red. If comparing the snapshots horizontally, it can be observed that larger values of $\eta$ (top) or $\nu$ (bottom) clearly promote the evolution of cooperation. More importantly, except for the left panel, $AC$ are always located in the center of the giant clusters, according to which we guess that this type of cooperators can promote the effective expansion of cooperator clusters. All the results are obtained for $b$=1.08.}
\end{center}
\end{figure}

\begin{figure}
\begin{center}
\includegraphics[width=9.5cm]{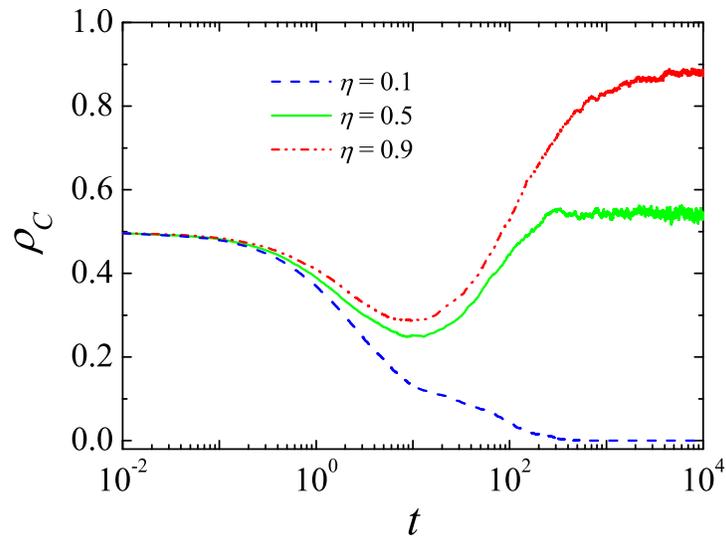}
\caption{Time courses of the fraction of cooperators $\rho_C$ independence on different values of $\eta$. Obviously, larger value of $\eta$ can stop the drop of cooperation at earlier stage and then turns it to the expansion of cooperation. The large the value of $eta$, the higher the final level of cooperation. During this process, the cooperators of type $A$ plays a crucial role for leading the effective expansion of clusters. All the results are obtained for $\nu=0.7$ and $b$=1.08.}
\end{center}
\end{figure}

\begin{figure}
\begin{center}
\includegraphics[width=13cm]{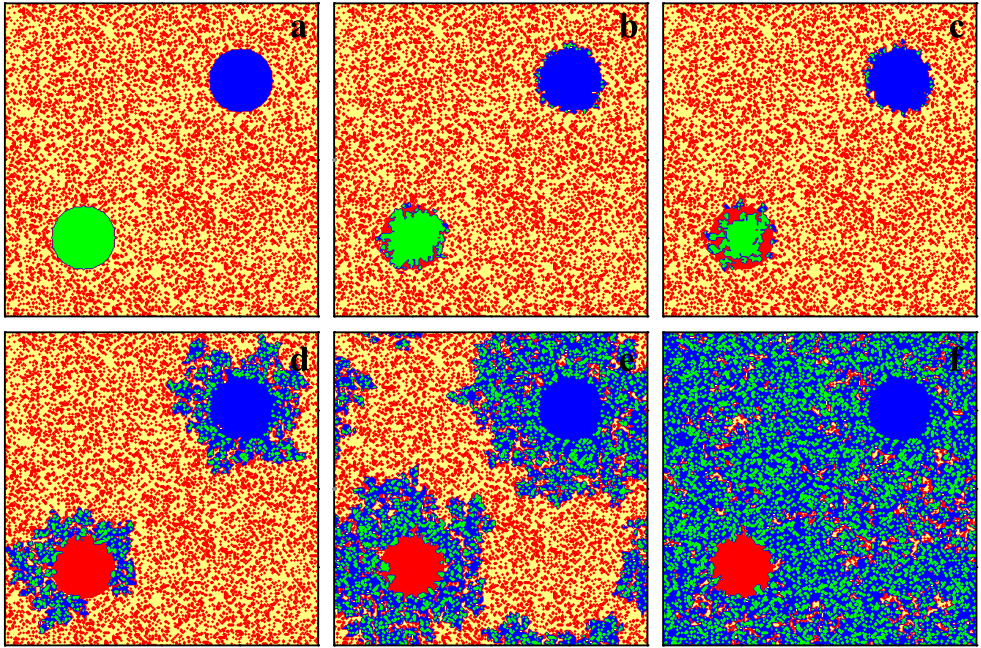}
\caption{Evolution snapshots of the clusters of cooperators on the square lattice. From {\bf a} to {\bf f}, the time steps are 0, 10, 50, 200, 1000 and 100000, respectively.
Different from previous treatments, there is a prepared initial state: two circular cooperative domains are present on both networks [one is for type $A$, ($AC$-blue), another domian is for type of $B$, ($BC$-green)], other agents are arranged as defectors. Except for blue domain, individuals of type $A$ are randomly chosen from defector ($AD$-yellow) to keep the value of $\eta$ accurate. This fig gives the direct observation for the leading role of $AC$, which can enable defectors to penetrate into the cooperator clusters. All the results are obtained for $\eta=0.75$, $\nu=0.75$ and $b$=1.08.}
\end{center}
\end{figure}

\end{document}